\documentclass[prd,showpacs,floatfix,twocolumn,nofootinbib,amsmath,amssymb,floatfix]{revtex4}
\usepackage{graphicx,color,dcolumn,booktabs,bm}
\usepackage{longtable,lscape}
\usepackage{txfonts}
\usepackage{overpic}
\usepackage{amssymb}
\usepackage{indentfirst}
\usepackage{feynmf}   %{feynmp}
\usepackage{slashed}  %for Feynman symbols
\usepackage{cases}
\usepackage{color}
\usepackage{multirow}
\usepackage{graphicx,color,dcolumn,booktabs,bm}

\graphicspath{{Figures/}} %
\graphicspath{{Figures/}} %
\usepackage[colorlinks, citecolor=blue,anchorcolor=red,menucolor=red, linkcolor=red,filecolor=red,runcolor=red,urlcolor=blue,frenchlinks=red]{hyperref}

\allowdisplaybreaks

\begin{document}
%\begin{CJK}{GBK}{}

%%%%%%%%%%%%%%%%%%%%%%%%%%%%%%%%%%%%%%%%%%%%%%%%%%%%%%%%%%%%%%%%%%%%%%%%%%%%%%%%%%%%%%%%%%%%%%%%%%%%%%%%%
%                                    The document begins here                                           %
%%%%%%%%%%%%%%%%%%%%%%%%%%%%%%%%%%%%%%%%%%%%%%%%%%%%%%%%%%%%%%%%%%%%%%%%%%%%%%%%%%%%%%%%%%%%%%%%%%%%%%%%%

\title{Is the newly reported $X(5568)$ a $B\bar{K}$ molecular state?}
\author{Rui Chen$^{1,2}$}\email{chenr2012@lzu.edu.cn}
\author{Xiang Liu$^{1,2}$\footnote{Corresponding author}}\email{xiangliu@lzu.edu.cn}
\affiliation{
$^1$School of Physical Science and Technology, Lanzhou University, Lanzhou 730000, China\\
$^2$Research Center for Hadron and CSR Physics, Lanzhou University and Institute of Modern Physics of CAS, Lanzhou 730000, China}

\date{\today}% It is always \today, today,

\begin{abstract}
In this work, we perform a dynamical study of the $B^{(*)}$ and $\bar K$ interaction and show that the newly reported $X(5568)$ or
$X(5616)$ cannot be assigned to be an isovector $B\bar{K}$ or $B^*\bar{K}$ molecular state. We continue to investigate the isoscalar $B^{(*)}\bar{K}$ systems, and the $B^{(*)}\bar{K}$ systems with isospin $I=0,1$, and
predict the existence of several isoscalar $B^{(*)}\bar K^{(*)}$ molecular states. A new task
of exploring open-bottom molecular states will be created for future experiments.

\end{abstract}

\maketitle

\section{introduction}\label{sec1}

In a recent experimental analysis \cite{D0:2016mwd}, the D\O\, Collaboration reported a new enhancement structure $X(5568)$ in the $B_s^0\pi^\pm$ invariant mass spectrum, which has mass $m=5567.8\pm2.9(\rm stat)^{+0.9}_{-1.9}(\rm syst)$ MeV and width $\Gamma=21.9\pm6.4(\rm stat)^{+5.0}_{-2.5}(\rm syst)$ MeV \cite{D0:2016mwd}. Due to its observed decay mode, we conclude that the $X(5568)$ must contain four different valence quark components, which makes the $X(5568)$ a good candidate for a tetraquark state. Experimental and theoretical exploration of exotic multiquark states has become an intriguing issue, especially with the experimental progress on charmonium-like $XYZ$ states and $P_c$ pentaquark states in the past 12 years (see the review papers \cite{Liu:2013waa,Chen:2016qju} for more details).

Before presenting the detailed analysis, we first focus on the concrete experimental information released by D\O\,\cite{D0:2016mwd}. The D\O\, measurement shows that the $X(5568)$ has spin-parity quantum number  $J^P=0^+$. However, there exists the possibility that the mass of the enhancement structure appearing in the $B_s^0\pi^\pm$ invariant mass spectrum would be shifted by the addition of the nominal mass difference $m_{B_s^*}-m_{B_s}$ \cite{D0:2016mwd}, which is due to the fact that the low-energy photons cannot be detected the in experiment. Thus, this enhancement structure may have a mass 5616 MeV, which corresponds to the $X(5616)$. Thus, the spin-parity of the $X(5616)$ is $J^P=1^+$ \cite{D0:2016mwd}.

Now that it has been observed $X(5568)$, theorists have paid more attention to the $X(5568)$.
The popular explanation of the $X(5568)$ as a tetraquark state composed of a diquark and antidiquark was proposed in Refs. \cite{Chen:2016mqt,Agaev:2016mjb,Wang:2016tsi,Zanetti:2016wjn,Wang:2016mee,Tang:2016pcf}.
In this interpretation, the decay $X(5568)\to B_s^0\pi^\pm$ was calculated using the QCD rum rule approach \cite{Agaev:2016ijz,Dias:2016dme,Wang:2016wkj}, which supports the $X(5568)$ as a tetraquark state. By making a calibration by the mass of the $X(5568)$, its partner states were predicted in Ref. \cite{Liu:2016ogz}, where the color-magnetic interaction was adopted and the tetraquark scenario was considered. In Ref. \cite{He:2016yhd}, He and Ko analyzed the symmetry properties of the $X(5568)$ and its partners based on flavor SU(3) symmetry.
Using a quark model with chromomagnetic interaction, the $X(5568)$ as a $su\bar{d}\bar{b}$ tetraquark was studied in Ref. \cite{Stancu:2016sfd}.
However, some groups hold opposite view. In a relativized quark model, the mass spectra of open-bottom tetraquark states were obtained \cite{Lu:2016zhe}. They found that the $X(5568)$ disfavors the assignment of the $sq\bar{b}\bar{q}$ tetraquark state since the theoretical result is higher than the data. In Ref. \cite{polosa}, Esposito {\it et al.} calculated the mass of the $X_b=[\bar b\bar q]_{S=0}[sq^\prime]_{S=0}$ state using the constituent quark model, which has the same quantum number as that of $X(5568)$. The mass of the $X(5568)$ is below the obtained mass of $X_b$.
Besides these tetraquark studies of the $X(5568)$, there were some discussions of the $X(5568)$ as the $ B\bar K$ molecular state  \cite{Xiao:2016mho,Agaev:2016urs}\footnote{There were some theoretical studies of the interactions between bottom-strange meson and kaon in Refs. \cite{Guo:2006fu,Guo:2006rp,Cleven:2010aw}. }. In Ref. \cite{Xiao:2016mho}, the $B_s^{*0}\pi^+$ decay width of the $X(5568)$ as the $ B\bar K$ molecular state was estimated, which is comparable with the experimental data on $X(5568)$. A QCD sum rule study in Ref. \cite{Agaev:2016urs} showed that a diquark-antidiquark configuration for the $X(5568)$ is more favorable than the $B \bar K$ molecular state picture.

In addition, the $X(5568)$ was explained to be the threshold effect \cite{Liu:2016xly}. We also noticed an investigation of the production of the $X(5568)$ in high-energy multiproduction process \cite{Jin:2016cpv}, where the authors indicated that it is hard to understand the large production rate of the $X(5568)$ using various general hadronization mechanisms. In recent work \cite{Burns:2016gvy,Guo:2016nhb}, the difficulty of explaining the $X(5568)$ as the $B\bar K$ molecular state
was indicated. The authors of Ref. \cite{Albaladejo:2016eps} further found that the $X(5568)$ signal can be reproduced by using $B_s\pi-B\bar{K}$ coupled channel analysis, if the corresponding cutoff value is larger than a natural value $\Lambda\sim 1$ GeV. Thus, they concluded that it is difficult to explain the properties of the $X(5568)$. Later, a further study along this line was given in Ref. \cite{Kang:2016zmv}.

When facing different proposals for the $X(5568)$, a crucial task is to find the evidence to distinguish these different explanations from the $X(5568)$. In this work, we perform a serious dynamical study of the interaction between $B^{(*)}$ and $\bar {K}$ using the one-boson exchange (OBE) model.
In this investigation, we check whether ${B}^{(*)}$ and $\bar K$ can be bound together to form a hadronic molecular state corresponding to the $X(5568)$ or the $X(5616)$.

This paper is organized as follows. We illustrate why the $X(5568)$ or the $X(5616)$ cannot be a $\bar{B}^{(*)}K$ molecular state in Sec. \ref{sec2} and Sec. \ref{sec3}. In Sec. \ref{sec4}, we present the prediction of the possible $\bar B^{(*)}K^{(*)}$ molecular states. Finally, the paper ends with a short summary.

\section{The $X(5568)$ cannot be an S-wave ${B}\bar K$ molecular state}\label{sec2}

The quantum number $I(J^P)$ for the $X(5568)$ is constrained as $1(0^+)$, since it has the decay channel $B_s^0\pi^{\pm}$. The flavor wave functions $|I,I_3\rangle$ of the ${B}\bar K$ system are defined as
$|1,1\rangle=|B^+\bar{K}^0\rangle$, $|1,0\rangle=\frac{1}{\sqrt{2}}\left(|B^+K^-\rangle-|B^0\bar{K}^0\rangle\right)$ and $|1,-1\rangle=|B^0{K}^-\rangle$. For the isoscalar ${B}\bar K$ system, its flavor wave function is $|0,0\rangle = \frac{1}{\sqrt{2}}\left(|B^+K^-\rangle+|B^0\bar{K}^0\rangle\right)$.
Here, we consider the S-wave $B\bar{K}$ molecular state \cite{Tornqvist:1993ng,Tornqvist:1993vu,Li:2012cs,Chen:2014mwa,Chen:2016heh}, which has the same quantum number as that of the $X(5568)$. Thus, the spin-orbit wave function of the $B\bar{K}$ system corresponds to $|{}^1S_0\rangle$ with spin $S=0$ and orbit $L=0$. In fact, we notice that the mass of the $X(5568)$ is about $206$ MeV lower than the $B\bar{K}$ threshold. This means that the $X(5568)$ should be a deeply bound state composed of ${B}$ and $\bar K$ if the $X(5568)$ is a $B\bar{K}$ molecular state. In the following,
we need to carry out a quantitative dynamical calculation to test this scenario.

In the OBE model, the interaction between ${B}$ and $\bar K$ can be due to the light vector-meson ($\rho$ and $\omega$) exchanges. The corresponding effective Lagrangians describing the couplings of $ B^{(*)} B^{(*)}\rho(\omega)$ \cite{Ding:2008gr,Sun:2011uh} and $\bar{K}^{(*)}\bar{K}^{(*)}\rho(\omega)$ \cite{Lin:1999ad} are
\begin{eqnarray}\label{eq:lag-v-exch}
  \mathcal{L}_{\widetilde{\mathcal{P}}^{(*)}\widetilde{\mathcal{P}}^{(*)}\mathbb{V}}
  &=& \sqrt{2}\beta{}g_V\widetilde{\mathcal{P}}^{\dag}_a
  \widetilde{\mathcal{P}}^{}_b
  v\cdot\mathbb{V}_{ab}
  -\sqrt{2}\beta g_V
  \widetilde{\mathcal{P}}^{*\dag}_a\cdot\widetilde{\mathcal{P}}_b^{*}
  v\cdot\mathbb{V}_{ab}\nonumber\\
  &&-i2\sqrt{2}\lambda{}g_V\widetilde{\mathcal{P}}^{*\mu\dag}_a
  \widetilde{\mathcal{P}}^{*\nu}_b\left(\partial_\mu{}
  \mathbb{V}_\nu - \partial_\nu{}\mathbb{V}_\mu\right)_{ab},\label{pspsv}\\
  \mathcal {L}_{\rho \widetilde{K}^{(*)}\widetilde{K}^{(*)}} &=& ig_{\rho \widetilde{K}\widetilde{K}}\left[\widetilde{K}^{\dag}\vec{\tau}\cdot \partial^{\mu}\widetilde{K}\vec{\rho_{\mu}}
       -\partial^{\mu}\widetilde{K}^{\dag}\vec{\tau}
       \cdot\widetilde{K}\vec{\rho_{\mu}}\right]\nonumber\\
       &&+ig_{\rho \widetilde{K}^*\widetilde{K}^*}\left[\left(\partial^{\mu}{\widetilde{K}^{*\nu\dag}}
       \widetilde{K}^*_{\nu}-{\widetilde{K}^{*\dag}_{\nu}}\partial^{\mu}
       \widetilde{K}^{*\nu}\right)
       \vec{\tau}\cdot\vec{\rho_{\mu}}\right.\nonumber\\
       &&+\left({\widetilde{K}^{*\dag}_{\mu}}\partial^{\mu}\widetilde{K}^{*\nu}
       -\partial^{\mu}{\widetilde{K}^{*\nu\dag}}\widetilde{K}^*_{\mu}\right)
       \vec{\tau}\cdot\vec{\rho_{\nu}}\nonumber\\
       &&\left.+\Big({\widetilde{K}^{*\dag}_{\nu}}\widetilde{K}^*_{\mu}
       -{\widetilde{K}^{*\dag}_{\mu}\widetilde{K}^*_{\nu}\Big)
       \vec{\tau}}\cdot\partial^{\mu}\vec{\rho^{\nu}}\right],\\
\mathcal {L}_{\omega \widetilde{K}^{(*)}\widetilde{K}^{(*)}} &=& ig_{\omega \widetilde{K}\widetilde{K}}\left[\widetilde{K}^{\dag}\partial^{\mu}
  \widetilde{K}\omega_{\mu}
       -\partial^{\mu}\widetilde{K}^{\dag}\widetilde{K}\omega_{\mu}\right]\nonumber\\
       &&+ig_{\omega \widetilde{K}^*\widetilde{K}^*}
       \left[\Big(\partial^{\mu}{\widetilde{K}^{*\nu\dag}}\widetilde{K}^*_{\nu}
       -{\widetilde{K}^{*\dag}_{\nu}}\partial^{\mu}
       \widetilde{K}^{*\nu}\Big)\omega_{\mu}\right.\nonumber\\
       &&+\Big({\widetilde{K}^{*\dag}_{\mu}}\partial^{\mu}\widetilde{K}^{*\nu}
       -{\widetilde{K}^{*\nu\dag}}
       \widetilde{K}^*_{\mu}\partial^{\mu}\Big)\omega_{\nu}\nonumber\\
       &&\left.+\Big({\widetilde{K}^{*\dag}_{\nu}}\widetilde{K}^*_{\mu}
       -{\widetilde{K}^{*\dag}_{\mu}}\widetilde{K}^*_{\nu}\Big)
       \partial^{\mu}\omega^{\nu}\right],\label{kk}
\end{eqnarray}
where the pseudoscalar $\widetilde{\mathcal{P}}$ and vector $\widetilde{\mathcal{P}}^*$ have the definition $\widetilde{\mathcal{P}}^{(*)}{}^T=\left(B^{(*)+}, {B}^{(*)0}, B_s^{(*)0}\right)$. The vector matrix $\mathbb{V}$ has the form
\begin{eqnarray}
\mathbb{V}&=&\left(\begin{array}{ccc}
\frac{\rho^{0}}{\sqrt{2}}+\frac{\omega}{\sqrt{2}} &\rho^{+}  &K^{*+}\\
\rho^{-}&-\frac{\rho^{0}}{\sqrt{2}}+\frac{\omega}{\sqrt{2}}  &K^{*0}\\
K^{*-}   &\bar{K}^{*0}  &\phi
\end{array}\right).
\end{eqnarray}
In addition, the coupling constants involved in Eq. (\ref{eq:lag-v-exch}) are taken as  $\beta=0.9$, $g_V=5.8$, and $\lambda=0.56 \,\text{GeV}^{-1}$ \cite{Sun:2011uh}, while the $KK\rho(\omega)$ constants $g_{\rho(\omega) K^{(*)}K^{(*)}}$ are
\begin{eqnarray}
g_{\rho \widetilde{K}^{(*)}\widetilde{K}^{(*)}} &=& -\frac{1}{4}g_1 = -3.425, \nonumber\\
g_{\omega \widetilde{K}^{(*)}\widetilde{K}^{(*)}} &=& -\frac{\sqrt{3}}{4}g_1\cos\theta = -4.396, \nonumber
\end{eqnarray}
which were given in Ref. \cite{Chen:2011cj}.

The effective potential of the isovector $B\bar{K}$ system is deduced as
\begin{eqnarray}
\mathcal{V}_{B\bar{K}}^{I=1} (r)&=&
      -\frac{\beta g_V}{2}\left[g_{\rho \widetilde{K}\widetilde{K}}Y(\Lambda,m_{\rho},r)
      -g_{\omega \widetilde{K}\widetilde{K}}Y(\Lambda,m_{\omega},r)\right].\label{bk1}
\end{eqnarray}
In the above expression, the cutoff factor $\Lambda$ denotes the phenomenological parameter around 1 GeV \cite{Tornqvist:1993ng,Tornqvist:1993vu}, which is introduced in the monopole form factor $\mathcal{F}(q^2,m_E^2)=({\Lambda^2-m_E^2})/({\Lambda^2-q^2})$ when writing out the scattering amplitude of $ B\bar K\to B\bar K$. Here, the function $Y(\Lambda,m,{r})$ reads as
\begin{eqnarray}
Y(\Lambda,m,{r}) &=&\frac{1}{4\pi r}(e^{-mr}-e^{-\Lambda
r})-\frac{\Lambda^2-m^2}{8\pi \Lambda}e^{-\Lambda r}.\label{yy}
\end{eqnarray}

\begin{figure}[htbp]
  \centering
  \includegraphics[width=0.46\textwidth]{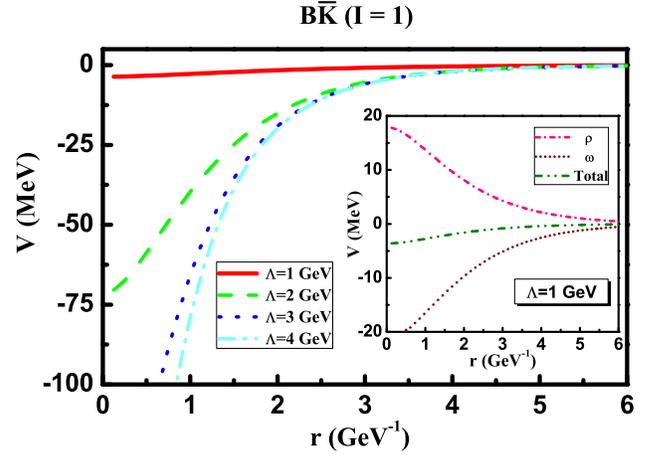}\\
\caption{The dependence of the OBE effective potential for the isovector S-wave $B\bar{K}$ system on $r$ and typical $\Lambda$ values. Here, we also show the variations of the subpotentials from the $\rho$ and $\omega$ meson exchanges to $r$.}\label{bki1}
\end{figure}

In Fig. \ref{bki1}, we first present the $r$ dependence of effective potentials for the isovector $B\bar{K}$ system, where we take several typical values of the cutoff $\Lambda$. As showed in Fig. \ref{bki1}, the total OBE effective potentials corresponding to $\Lambda=1 \sim 4$ GeV are attractive. As the values of $\Lambda$ increases, the attraction between ${B}$ and $\bar K$ becomes stronger.
Furthermore, we numerically solved the Schr\"odinger equation with the obtained effective potential, and could not find the corresponding bound-state solution for this S-wave isovector $B\bar{K}$ system when taking $\Lambda=1 \sim 5$ GeV \cite{Tornqvist:1993ng,Tornqvist:1993vu}, which means that the ${B}$ and $\bar K$ cannot be bound together to form an S-wave $B\bar{K}$ molecular state with isospin $I=1$.

{Since the $X(5568)$ was observed in the $B_s^+\pi^0$ channel, which is close to the mass of the $X(5568)$,
we further consider the coupled-channel effect due to the mixing between the $B_s^+\pi^0$ and $B^+\bar{K}^0$ channels. In our calculation, we adopt the effective potential {\cite{Lin:1999ad}}
\begin{eqnarray}
\mathcal {L}_{\pi \widetilde{K}\widetilde{K}^*} &=&
      ig_{\pi\widetilde{K}\widetilde{K}^*}
      \left[\widetilde{K}^{\dag}\vec{\tau}\cdot \widetilde{K}^{*\mu}\partial_{\mu}\vec{\pi}
      -\widetilde{K}^{\dag}\vec{\tau}\cdot\partial_{\mu}
      \widetilde{K}^{*\mu}\vec{\pi}\right]+H.c.,
\end{eqnarray}
{where $g_{\pi\widetilde{K}\widetilde{K}^*}=\frac{1}{4}g_1$ \cite{Chen:2011cj}. }
Then, the obtained total effective potentials corresponding to the discussed $X(5568)$ can be written as
\begin{eqnarray}
\mathcal{V}(r) &=& \left(\begin{array}{cc}
     \langle B_s\pi|V|B_s\pi\rangle  &\langle B_s\pi|V|B\bar{K}\rangle\\
     \langle B\bar{K}|V|B_s\pi\rangle &\langle B\bar{K}|V|B\bar{K}\rangle
     \end{array}\right)
\end{eqnarray}
with
\begin{eqnarray*}
\langle B_s\pi|V|B_s\pi\rangle &=& 0,\\
\langle B_s\pi|V|B\bar{K}\rangle &=& \langle B\bar{K}|V|B_s\pi\rangle\nonumber\\
    &=&\frac{\sqrt{2}}{4}\beta g_Vg_{\pi\widetilde{K}\widetilde{K}^*}
    \left(m_{\pi}+m_K\right)Y(\Lambda,m_{K^*},r),\\
\langle B\bar{K}|V|B\bar{K}\rangle &=& -\frac{\beta g_V}{2}\left[g_{\rho \widetilde{K}\widetilde{K}}Y(\Lambda,m_{\rho},r)
      -g_{\omega \widetilde{K}\widetilde{K}}Y(\Lambda,m_{\omega},r)\right].\nonumber\\
\end{eqnarray*}
With this deduced effective potential, we solve the coupled-channel Schr\"odinger equation. Unfortunately,
we still cannot find the bound-state solutions when scanning the range $\Lambda=1\sim 5$ GeV.

According to our study, we can fully exclude the $X(5568)$ as an isovector S-wave $B\bar{K}$ molecular state with $J^P=0^+$, which is consistent with the conclusion made in Refs. \cite{Zhang:2006ix,Liu:2009uz}.

\section{The $X(5616)$ cannot be an S-wave ${B}^*\bar K$ molecular state}\label{sec3}

Since the quantum number $I(J^P)$ of the $X(5616)$ is $1(1^+)$ \cite{D0:2016mwd}, the S-wave ${B}^*\bar K$ molecular state is possible assignment for the $X(5616)$. If we only consider the S-wave interaction between ${B}^*$ and $\bar K$ mesons, the obtained OBE effective potential is
\begin{eqnarray}
\mathcal{V}_{{B}^*\bar K}^{I=1}(r) &=&
      -\frac{\beta g_V}{2}\left[g_{\rho \widetilde{K}\widetilde{K}}
      Y(\Lambda,m_{\rho},r)-g_{\omega \widetilde{K}\widetilde{K}}
      Y(\Lambda,m_{\omega},r)\right],\label{bsk1}\quad\,\,
\end{eqnarray}
which is the same as the expression in Eq. (\ref{bk1}). The difference between $B\bar K$ and ${B}^*\bar K$ with $I=1$ can be seen in the difference of their reduced masses.
Although the total effective potential of an S-wave ${B}^*\bar K$ system with isospin $I=1$ is attractive, we cannot find the corresponding bound-state solution.

When further considering the S-D mixing effect on the ${B}^*\bar K$ system since {there exists mixing of the $B^*\bar K$ systems with spin-orbit wave functions $|{}^3S_1\rangle$ and $|{}^3D_1\rangle$}, the effective potential in Eq. (\ref{bsk1}) should be modified as
\begin{eqnarray}
\mathcal{V}_{{B}^*\bar K}^{I=1}(r) &=&
      -\frac{\beta g_V}{2}
      \left(\begin{array}{cc}
      1   &0\\
      0   &1
      \end{array}\right)
      \left[g_{\rho \widetilde{K}\widetilde{K}}
      Y(\Lambda,m_{\rho},r)\right.\nonumber\\&&\left.-g_{\omega \widetilde{K}\widetilde{K}}
      Y(\Lambda,m_{\omega},r)\right],\label{bsk2}
\end{eqnarray}
which is a $2\times 2$ matrix, where { the matrix ${\text{diag}}(1,1)$ is deduced from
\begin{eqnarray}
\left(\begin{array}{cc}
\langle{}^3S_1|{\bf{\epsilon_1\cdot\epsilon_3^{\dag}}}|{}^3S_1\rangle
&\langle{}^3S_1|{\bf{\epsilon_1\cdot\epsilon_3^{\dag}}}|{}^3D_1\rangle\\
\langle{}^3D_1|{\bf{\epsilon_1\cdot\epsilon_3^{\dag}}}|{}^3S_1\rangle
&\langle{}^3D_1|{\bf{\epsilon_1\cdot\epsilon_3^{\dag}}}|{}^3D_1\rangle
\end{array}\right)=\left(\begin{array}{cc}
      1   &0\\0   &1
\end{array}\right).
\end{eqnarray}
Here, $\epsilon_1$ and $\epsilon_3^{\dag}$ correspond to the operators of the polarization vectors of the initial and finial $B^*$ meson, respectively.} To search for the bound-state solution, we solve the coupled-channel Schr\"odinger equation with Eq. (\ref{bsk2}). The bound-state solution is still absent when we scan the range $\Lambda=1\sim 5$ GeV in our numerical analysis.

{ In our calculation, we further consider the coupled-channel effect with the $B_s^*\pi$ and $B^*\bar{K}$ channels. However, the bound solutions cannot obtained.
}

Thus, our study does not support the $X(5616)$ as an isovector S-wave ${B}^*\bar K$ molecular state.

\section{The prediction of possible ${B}^{(*)}\bar K^{(*)}$ molecular states}\label{sec4}

\subsection{Isoscalar ${B}\bar K$ and ${B}^*\bar K$ systems}\label{sec41}
In the above sections, we discussed isovector ${B}\bar K$ and ${B}^*\bar K$ systems, which also stimulates our interest in further studying other ${B}^{(*)}\bar K^{(*)}$ systems. First, we focus on the isoscalar ${B}\bar K$ and ${B}^*\bar K$ systems. Their OBE effective potentials are
\begin{eqnarray}
\mathcal{V}_{{B}\bar K}^{I=0} (r)&=&
      \frac{\beta g_V}{2}\left[3g_{\rho \widetilde{K}\widetilde{K}}Y(\Lambda,m_{\rho},r)
     +g_{\omega \widetilde{K}\widetilde{K}}Y(\Lambda,m_{\omega},r)\right],\\
\mathcal{V}_{{B}^*\bar K}^{I=0}(r) &=&
      \frac{\beta g_V}{2}
      \left(\begin{array}{cc}
      1   &0\\
      0   &1
      \end{array}\right)
      \left[3g_{\rho \widetilde{K}\widetilde{K}}
      Y(\Lambda,m_{\rho},r)\right.\nonumber\\&&\left.+g_{\omega \widetilde{K}\widetilde{K}}
      Y(\Lambda,m_{\omega},r)\right].\label{bsk0}
\end{eqnarray}
When comparing the OBE effective potentials of the isoscalar and isovector ${B}^{(*)}\bar K$ systems, we find that an isospin factor $-3$ is introduced in the $\rho$-exchange potentials for these isoscalar systems, while the isoscalar and isovector ${B}^{(*)}\bar K$ systems have the same $\omega$-exchange potential. The behaviors of the effective potentials of the isoscalar ${B}^{(*)}\bar K$ systems make that it easier to form the isoscalar ${B}^{(*)}\bar K$ molecular states.
By solving the Schr\"odinger equation, we confirm the above speculation, namely that we can find the bound-state solutions for the  isoscalar ${B}^{(*)}\bar K$ systems. In Table. \ref{num0}, we list the obtained binding energy, root-mean-square radius and the corresponding $\Lambda$ values. When taking $\Lambda=1.9$ GeV, there exist shallow isoscalar ${B}^{(*)}\bar K$ molecular states.As the value of $\Lambda$ increases, the binding energies of these two systems become deeper. Here, the input of $\Lambda$ is not far away from 1 GeV, which come from studying the nuclear force \cite{Tornqvist:1993ng,Tornqvist:1993vu}. Thus, we may conclude that there probably exist isoscalar ${B}\bar K$ and ${B}^{*}\bar K$ molecular states, which have the quantum numbers $I(J^P)=0(0^+)$ and $I(J^P)=0(1^+)$, respectively.

\renewcommand{\arraystretch}{1.5}
\begin{table}[!hbtp]
\caption{The $\Lambda$ dependence of the obtained bound-state solutions (binding energy $E$ and root-mean-square radius $r_{RMS}$) for isoscalar ${B}^{(*)}\bar K$ systems. Here, $E$, $r_{RMS}$, and $\Lambda$ are in units of MeV, fm, and GeV, respectively.}\label{num0}
\begin{tabular}{ccccccccc}
\toprule[1pt]\toprule[1pt]
State &$\Lambda$   &$E$ &$r_{RMS}$     &\quad\quad
 &State  &$\Lambda$    &$E$  &$r_{RMS}$ \\ \midrule[1pt]
 $[{B}\bar K]_{J=0}^{I=0}$    &1.90    &-0.29    &5.66    &
 &$[{B}^*\bar K]_{J=1}^{I=0}$      &1.90      &-0.30    &5.64\\
                             &2.10    &-4.36    &2.45    &
                                 & &2.10      &-4.40    &2.44\\
                             &2.30    &-11.69   &1.58    &
                                 & &2.30      &-11.76   &1.57\\
 \bottomrule[1pt]
\bottomrule[1pt]
\end{tabular}
\end{table}

{In fact, the above formula can be extended to the discussion of the
$DK$ system with $(I=0, J=0)$ and the $D^*K$ system with $(I=0, J=1)$.
Our calculation shows that the masses of the $D_{s0}(2317)$ and the $D_{s1}^*(2460)$ \cite{Agashe:2014kda} can be reproduced when the cutoff $\Lambda$ is taken around 3.5 GeV, where the $D_{s0}(2317)$ and the $D_{s1}^*(2460)$ correspond to the
$DK$ system with $(I=0, J=0)$ and the $D^*K$ system with $(I=0, J=1)$, respectively, since the reduced masses of the $BK$ and $B^*\bar{K}$ systems are heavier than those of the $DK$ and $D^*K$ systems, respectively. Thus, we can conclude that the cutoff $\Lambda$ for $BK/B^*\bar{K}$ should be smaller than that of
$DK/D^*K$. The numerical results listed in Table \ref{num0} indeed can reflect this point. }

If isoscalar ${B}\bar K$ and ${B}^{*}\bar K$ molecular states exist, finding them becomes a crucial task. For an isoscalar ${B}\bar K$ molecular state, its two-body and three-body Okubo-Zweig-Iizuka-allowed decay channels are forbidden. Thus, experimental searches for this isoscalar ${B}\bar K$ are very difficult. For an isoscalar ${B}^{*}\bar K$ molecular state, we suggest an experiment to further analyze its $B_s\pi\pi$ final state, by which this isoscalar ${B}^{*}\bar K$ molecular state can be discovered.

\subsection{The $B\bar K^*$ and $B^*\bar K^*$ systems}

Besides the systems discussed in Sec. \ref{sec2} and \ref{sec41}, in this work we also investigate the $B\bar K^*$ and $B^*\bar K^*$ systems. For the $B^*\bar K^*$ systems, there also exist $\pi$ and $\eta$ meson-exchange contributions to the effective potentials. In deducing the effective potentials, we need to adopt the following effective Lagrangians:
\begin{eqnarray}
\mathcal{L}_{\widetilde{\mathcal{P}}^*\widetilde{\mathcal{P}}^*\mathbb{P}} &=&
     i \frac{2g}{f_\pi}\varepsilon_{\alpha\mu\nu\lambda}v^\alpha
     \widetilde{\mathcal{P}}^{*\mu\dag}_{a}
     \widetilde{\mathcal{P}}^{*\lambda}_{b}\partial^\nu{}\mathbb{P}_{ab},\label{ppp}\\
\mathcal {L}_{\pi \widetilde{K}^*\widetilde{K}^*} &=&
        -g_{\pi \widetilde{K}^*\widetilde{K}^*}
        \varepsilon^{\mu\nu\rho\sigma}
        \partial_{\rho}{\widetilde{K}^{*\dag}_{\sigma}}
        \vec{\tau}\cdot\partial_{\mu}\widetilde{K}^*_{\nu}\vec{\pi},\label{kkp}\\
\mathcal {L}_{\eta \widetilde{K}^*\widetilde{K}^*} &=&
        g_{\eta \widetilde{K}^*\widetilde{K}^*}\varepsilon^{\mu\nu\rho\sigma}
        \partial_{\rho}{\widetilde{K}^{*\dag}_{\sigma}}\partial_{\mu}
        \widetilde{K}^*_{\nu}\eta\label{kke}
\end{eqnarray}
with
\begin{eqnarray}
    {\mathbb P}&=&\left(\begin{array}{ccc}
        \frac{\pi^0}{\sqrt{2}}+\frac{\eta}{\sqrt{6}}&\pi^+  &K^+\\
        \pi^-&-\frac{\pi^0}{\sqrt{2}}+\frac{\eta}{\sqrt{6}}  &K^0\\
        K^-  &\bar{K}^0   &-\frac{2\eta}{\sqrt{6}}
\end{array}\right).
\end{eqnarray}
Here, $g=0.59$ is extracted from the experimental width of $D^{*+}$ \cite{Isola:2003fh}, and the pion decay constant $f_{\pi}=132$ MeV. Additionally, $g_{\pi \widetilde{K}^*\widetilde{K}^*}$ and $g_{\eta \widetilde{K}^*\widetilde{K}^*}$ are expressed by $g_{\pi \widetilde{K}^*\widetilde{K}^*}  =\frac{g_1^2N_c}{64\pi^2f_{\pi}}$, and
$g_{\eta \widetilde{K}^*\widetilde{K}^*}   =\frac{g_1^2N_c}{64\sqrt{3}\pi^2f_{\pi}}$ \cite{Kaymakcalan:1983qq}
 with the number of colors $N_c$, where the value of $g_1$ was given in Sec. \ref{sec2}.

{ Here, the S-D mixing effect is also taken into account, and the relevant spin-orbit wave functions $|{}^{2S+1}L_J\rangle$ include
\begin{eqnarray}
\begin{array}{ccccc}
B\bar{K}^*:    &|{}^3S_1\rangle,  &|{}^3D_1\rangle,\\
B^*\bar{K}^*:  &|{}^1S_0\rangle,  &|{}^5D_0\rangle,\\
               &|{}^3S_1\rangle,  &|{}^3D_1\rangle,  &|{}^5D_1\rangle,\\
               &|{}^5S_2\rangle,  &|{}^1D_2\rangle,  &|{}^3D_2\rangle,  &|{}^5D_2\rangle.
\end{array}
\end{eqnarray}}

The obtained general expressions of the ${B}\bar K^*$ and ${B}\bar K^*$ systems when considering the S-D mixing effect read
\begin{eqnarray}
\mathcal{V}^{I}_{{B}\bar K^*}(r) &=& \frac{1}{2}\mathcal{G}(I)\beta g_Vg_{\rho \widetilde{K}^*\widetilde{K}^*}
      \left(\begin{array}{cc}
      1   &0\\
      0   &1
      \end{array}\right)Y(\Lambda,m_{\rho},r)\nonumber\\
      &&+\frac{1}{{2}}\beta g_Vg_{\omega \widetilde{K}^*\widetilde{K}^*}
      \left(\begin{array}{cc}
      1   &0\\
      0   &1
      \end{array}\right)Y(\Lambda,m_{\omega},r),\\
{\mathcal{V}}^{I,J}_{{B}^*\bar K^*} (r)&=& \frac{1}{6\sqrt{2}}
      \frac{gg_{\pi \widetilde{K}^*\widetilde{K}^*}}{f_{\pi}}\mathcal{G}(I)\Bigg[\mathcal{E}_1(J)\nabla^2+\mathcal{S}(J)
      r\frac{\partial}{\partial r}\frac{1}{r}\frac{\partial}{\partial r}\Bigg]\nonumber\\&&\times
      Y(\Lambda,m_{\pi},r)-\frac{1}{6\sqrt{6}}\frac{gg_{\eta \widetilde{K}^*\widetilde{K}^*}}{f_{\pi}}\Bigg[
      \mathcal{E}_1(J)\nabla^2\nonumber\\&&+\mathcal{S}(J)
      r\frac{\partial}{\partial r}\frac{1}{r}\frac{\partial}{\partial r}\Bigg]
      Y(\Lambda,m_{\eta},r)\nonumber\\
      &&-\frac{1}{2}\beta g_Vg_{\rho \widetilde{K}^*\widetilde{K}^*}\mathcal{G}(I)
      \mathcal{E}_2(J)Y(\Lambda,m_{\rho},r),\nonumber\\
      &&+\frac{1}{2}\beta g_Vg_{\omega \widetilde{K}^*\widetilde{K}^*}
      \mathcal{E}_2(J)Y(\Lambda,m_{\omega},r),\label{bsks}
\end{eqnarray}
where the superscripts $I$ and $J$ denote the isospin and total angular momentum of these discussed systems.
$\mathcal{G}(I)$ is the isospin factor, which is taken as $-3$ for the isoscalar system, and $1$ for the isovector system.
The concrete forms of $\mathcal{E}_1(J)$, $\mathcal{E}_2(J)$, and $\mathcal{S}(J)$ are  $\mathcal{E}_1(0)=\text{diag}(2,-1)$, $\mathcal{E}_1(1)=\text{diag}(1,1,-1)$, $\mathcal{E}_1(2)=\text{diag}(-1,2,1,-1)$,
$\mathcal{E}_2(0)=\text{diag}(1,1)$, $\mathcal{E}_2(1)=\text{diag}(1,1,1)$, $\mathcal{E}_2(2)=\text{diag}(1,1,1,1)$, $\mathcal{S}(0)=\left(\begin{array}{cc}0 &\sqrt{2}\\  \sqrt{2}  &2\end{array}\right)$, $\mathcal{S}(1)=\left(\begin{array}{ccc}0 &-\sqrt{2} &0\\  -\sqrt{2}  &1   &0\\  0  &0  &1\end{array}\right)$, and $\mathcal{S}(2)=\tiny{\left(\begin{array}{cccc} 0 &\sqrt{\frac{2}{5}} &0   &-\sqrt{\frac{14}{5}}\\  \sqrt{\frac{2}{5}}  &0   &0   &-\frac{2}{\sqrt{7}}\\  0  &0  &-1  &0\\  -\sqrt{\frac{14}{5}}  &-\frac{2}{\sqrt{7}}   &0  &-\frac{3}{7}\end{array}\right)}$.

\renewcommand{\arraystretch}{1.5}
\begin{table}[!hbtp]
\caption{The $\Lambda$ dependence of the obtained bound-state solutions (binding energy $E$ and root-mean-square radius $r_{RMS}$) of the $B\bar K^*$ and $B^*\bar K^*$ systems. Here, $E$, $r_{RMS}$, and $\Lambda$ are in units of MeV, fm, and GeV, respectively.}\label{num}
\begin{tabular}{ccccccccc}
\toprule[1pt]\toprule[1pt]
State &$\Lambda$   &$E$ &$r_{RMS}$     &\quad\quad
 &State  &$\Lambda$    &$E$  &$r_{RMS}$ \\ \midrule[1pt]

 $[{B}\bar K^*]_{J=1}^{I=0}$   &1.40   &-0.32    &5.16    &
 &$[{B}\bar K^*]_{J=1}^{I=1}$       &\ldots   &\ldots   &\ldots\\
                              &1.60   &-10.30   &1.37
                              &&&\ldots   &\ldots   &\ldots\\
                              &1.80   &-30.20   &0.88
                              &&&\ldots   &\ldots   &\ldots\\
 $[{B}^*\bar K^*]_{J=0}^{I=0}$    &0.88   &-0.60    &4.91 &
 &$[{B}^*\bar K^*]_{J=0}^{I=1}$          &3.00    &-0.98     &3.67\\
                                 &1.08   &-6.06    &2.04
                                       &&&3.30   &-6.57     &1.55\\
                                 &1.28   &-20.97    &1.24
                                       &&&3.60   &-19.34    &0.94\\
 $[{B}^*\bar K^*]_{J=1}^{I=0}$   &1.60   &-1.15    &3.62    &
 &$[{B}^*\bar K^*]_{J=1}^{I=1}$          &\ldots   &\ldots   &\ldots\\
                                &1.80    &-8.69    &1.54
                                &&&\ldots   &\ldots   &\ldots\\
                                &2.00    &-22.40   &1.06
                                &&&\ldots   &\ldots   &\ldots\\
 $[{B}^*\bar K^*]_{J=2}^{I=0}$   &1.10   &-0.14    &5.77    &
 &$[{B}^*\bar K^*]_{J=2}^{I=1}$          &\ldots   &\ldots   &\ldots\\
                                &1.20    &-7.41    &1.57
                                &&&\ldots   &\ldots   &\ldots\\
                                &1.30    &-24.48   &0.97
                                &&&\ldots   &\ldots   &\ldots\\
\bottomrule[1pt]
\bottomrule[1pt]
\end{tabular}
\end{table}

With the above preparation, we try to search for the bound solutions by solving the Schr$\ddot{\text{o}}$dinger equation. In Table \ref{num}, the obtained results are collected. {Among the discussed isovector $B\bar K^*$ and $B^*\bar K^*$ systems, only the $B^*\bar{K}^*$ system with $J=0$ has a bound-state solution when $\Lambda$ is around 3 GeV, which is obviously different from 1 GeV \cite{Tornqvist:1993ng,Tornqvist:1993vu}. Thus, if strictly considering this criterion of the $\Lambda$ value, we conclude that there do not exist isovector $B^{(*)}\bar K^*$ molecular states. }Different from the isovector case, the isoscalar $B^{(*)}\bar K$ systems may exist, as shown in Table \ref{num}.
In the following, we further discuss their allowed decay modes:
\begin{enumerate}
\item The ${B}\bar K^*$ molecular state with $(I=0,J=1)$ can decay into $B^*\bar{K}$, ${B}_s\omega$ and ${B}_s^*\eta$.
\item ${B}_s^*\omega$ is an allowed decay mode of the ${B}^*\bar K^*$ molecular state with $(I=0,J=2)$.
\item The allowed decay channels of the ${B}^*\bar K^*$ molecular state with $(I=0,J=1)$ include ${B}^*\bar K$, ${B}\bar K^*$, ${B}_s\omega$, and ${B}_s^*\omega$.
\item ${B}\bar K$, ${B}_s\eta$ and ${B}_s^*\omega$ are the allowed two-body decay channels for the ${B}^*\bar K^*$ state with $(I=0,J=0)$.
\end{enumerate}

{ In our calculation, we also extend our study to the charm sector. The relevant numerical results for the $DK^*$ and $D^*K^*$ systems are collected in Table \ref{pre}.

\renewcommand{\arraystretch}{1.5}
\begin{table}[!hbtp]
\caption{The $\Lambda$ dependence of the obtained bound-state solutions (binding energy $E$ and root-mean-square radius $r_{RMS}$) of the $DK^*$ and $D^*K^*$ systems. Here, $E$, $r_{RMS}$, and $\Lambda$ are in units of MeV, fm, and GeV, respectively.}\label{pre}
\begin{tabular}{ccccccccc}
\toprule[1pt]\toprule[1pt]
State &$\Lambda$   &$E$ &$r_{RMS}$     &\quad\quad
 &State  &$\Lambda$    &$E$  &$r_{RMS}$ \\ \midrule[1pt]

 $[{D} K^*]_{J=1}^{I=0}$   &1.60   &-0.90    &4.18    &
 &$[{D} K^*]_{J=1}^{I=1}$       &\ldots   &\ldots   &\ldots\\
                              &1.80   &-9.30   &1.56
                              &&&\ldots   &\ldots   &\ldots\\
                              &2.00   &-23.87   &1.05
                              &&&\ldots   &\ldots   &\ldots\\
 $[{D}^*K^*]_{J=0}^{I=0}$    &1.00   &-0.79    &4.76 &
 &$[{D}^*K^*]_{J=0}^{I=1}$          &3.70    &-0.46     &4.92\\
                                 &1.20   &-6.97    &2.05
                                       &&&4.10   &-7.88     &1.56\\
                                 &1.40   &-22.51    &1.27
                                       &&&4.50   &-28.87    &0.85\\
 $[{D}^*K^*]_{J=1}^{I=0}$   &1.80   &-0.89    &4.25    &
 &$[{D}^*K^*]_{J=1}^{I=1}$          &\ldots   &\ldots   &\ldots\\
                                &2.20    &-15.92    &1.29
                                &&&\ldots   &\ldots   &\ldots\\
                                &2.60    &-47.17   &0.84
                                &&&\ldots   &\ldots   &\ldots\\
 $[{D}^*K^*]_{J=2}^{I=0}$   &1.20   &-0.21    &5.66    &
 &$[{D}^*K^*]_{J=2}^{I=1}$          &\ldots   &\ldots   &\ldots\\
                                &1.30    &-6.80    &1.77
                                &&&\ldots   &\ldots   &\ldots\\
                                &1.40    &-21.52   &1.09
                                &&&\ldots   &\ldots   &\ldots\\
\bottomrule[1pt]
\bottomrule[1pt]
\end{tabular}
\end{table}
These numerical results shown in Table \ref{pre} indicate that the isoscalar $DK^*$ and $D^*K^*$ states are very promising molecular candidates. Their decay behaviors are
\begin{eqnarray}
&&[DK^*]_{J=1}^{I=0} \to D^*K, D_s\eta, D_s^*\omega,
          \nonumber\\
&&[D^*K^*]_{J=1}^{I=0}
\to DK, D_s\eta, D_s^*\omega,\nonumber\\
&&[D^*K^*]_{J=1}^{I=0}\to D^*K, DK^*, D_s\omega, D_s^*\omega,
    \nonumber\\
&&[D^*K^*]_{J=2}^{I=0}\to D_s^*\omega.\nonumber
\end{eqnarray}
}

It is obvious that experimental searches for these predicted isoscalar $B^{(*)}\bar K^*$ and $D^{(*)}K^*$ molecular states will be an intriguing issue.
The above information is valuable to further study them experimentally.

\section{Summary}\label{sec5}

Stimulated by the recent evidence of a new enhancement structure $X(5568)$ or $X(5616)$ \cite{D0:2016mwd},
we carried out a study of the interactions of isovector $B\bar{K}$ and $B^*\bar{K}$ systems via the OBE model. This dynamical study makes us exclude the $X(5568)$ or the $X(5616)$ as the isovector $B\bar{K}$ or $B^*\bar{K}$ molecular state.
In Refs. \cite{Burns:2016gvy,Guo:2016nhb}, the difficulty of assigning the $X(5568)$ to be the $B\bar K$ molecular state was discussed. Obviously,  we reach the same conclusion using different approaches.

In this work, we also studied isoscalar $B\bar{K}$ and $B^*\bar{K}$ systems; we predicted that there isoscalar $B\bar{K}$ and $B^*\bar{K}$ molecular states may exist, and their decay behaviors were discussed. In addition, we also focused on the $B^{(*)}\bar{K}^*$ systems. Our calculation illustrates that $B^{(*)}$ and $\bar{K}^*$ cannot form isovector molecular states, but they can be bound together to construct isoscalar $B^{(*)}\bar{K}^*$ molecular states. The allowed decay modes of these possible isoscalar $B^{(*)}\bar{K}^*$ molecular states show that it is possible to find them in experiments. Thus, we suggest future experimental exploration of these isoscalar open-bottom molecular states.

\subsection*{Acknowledgments}
This project is supported by
the National Natural Science Foundation of China under Grants No. 11222547 and No. 11175073 and the Fundamental Research Funds for the Central Universities. X. L. is also supported by the National Youth Top-notch Talent Support Program (Thousands-of-Talents Scheme).

{\it Note added}-When preparing the manuscript,
we noticed the preliminary result from the LHCb experiment \cite{lhcb}, where the signal of $X(5568)$ was not observed.
In Ref. \cite{lhcb}, the LHCb's analysis also
shows that the cone cut selection criterion can generate broad peaking
structures. The D\O\, Collaboration
performed an analysis of the $B_s^0\pi^+$ data with and without the cone cut, which indicates that there exists a structure with and without the cone cut.
Here, the cone cut clearly enhances the resonance state as analyzed
in Ref. \cite{D0:2016mwd}. According to our present study, we can deny the possibility of the $X(5568)$ or $X(5616)$ as an isoscalar $B\bar K$ or $B^*K$ hadronic molecular state.

\end{document}